\documentclass[twocolumn,a4paper]{article}

\usepackage{graphicx}          

\usepackage{amsmath}
\usepackage{amssymb}

\usepackage{mathrsfs}       

\def\EE{\mathbb{E}}
\def\PP{\mathbb{P}}
\def\ZZ{\mathbb{Z}}
\def\NN{\mathbb{N}}
\def\RR{\mathbb{R}}
\def\CC{\mathbb{C}}

\def\T{\mathscr{T}}

\def\Q{\mathscr{Q}}

\def\R{\mathscr{R}}
\def\V{\mathscr{V}}
\def\I{\mathscr{I}}
\def\T{\mathscr{T}}
\def\Lscr{\mathscr{L}}
\def\S{\mathscr{S}}
\def\F{\mathscr{F}}

\def\C{\mathscr{C}}

\def\diag{\text{diag}}

\def\var{\text{var}}
\def\std{\text{std}}
\def\mean{\text{mean}}

\def\Var{\text{Var}}
\def\Cov{\text{Cov}}

\def\mRNA{\textit{m}RNA }

\def\conv{*}

\def\newt{z} 
\def\sl{s}

\def\vind{I}
\usepackage{bbm}
\def\ind{\mathbbm{1}}

\usepackage{subfigure}

\newtheorem{proposition}{Proposition}

\newtheorem{corollary}{Corollary}

\title{
Stochastic reaction networks with input processes: \\ Analysis and applications to reporter gene systems
}
\author{Eugenio Cinquemani\,\thanks{Inria Grenoble -- Rh\^one-Alpes, 655 Avenue de l'Europe, Montbonnot, 38334 Saint-Ismier cedex, France. \newline \textit{E-mail:} \texttt{eugenio.cinquemani@inria.fr}}}

\date{}

\begin{document}

\maketitle

\begin{abstract}                        
Stochastic reaction network models are widely utilized in biology and chemistry to describe the probabilistic dynamics of biochemical systems in general, and gene interaction networks in particular. Most often, statistical analysis and inference of these systems is addressed by parametric approaches, where the laws governing exogenous input processes, if present, are themselves fixed in advance. Motivated by reporter gene systems, widely utilized in biology to monitor gene activation at the individual cell level, we address the analysis of reaction networks with state-affine reaction rates and arbitrary input processes. We derive a generalization of the so-called moment equations where the dynamics of the network statistics are expressed as a function of the input process statistics. In stationary conditions, we provide a spectral analysis of the system and elaborate on connections with linear filtering. We then apply the theoretical results to develop a method for the reconstruction of input process statistics, namely the gene activation autocovariance function, from reporter gene population snapshot data, and demonstrate its performance on a simulated case study.
\end{abstract}

\section{Introduction}

At the level of individual molecules, biochemical reaction network dynamics are determined by random encounters of molecules of the different participating species. Under suitable assumptions on the reaction volume, the stochastic dynamics of the network are most often described in terms of Continuous-Time Markov Chains (CTMC) where the abundance of the different species constitutes the random system state, and determines the instantaneous propensity of the different reactions~\cite{Gillespie1992}. Stochastic reaction network modelling is widely utilized in nowadays's research in biology, in particular, to analyze and understand gene expression dynamics and interactions~\cite{Raj2008,OzbudakEtAl2002,ThattaiVanOudenaarden2001,KaernEtAl2005,Paulsson2005157}. Correspondingly, tools such as the Chemical Master Equation (CME,~\cite{Gillespie1992}) and the Moment Equations (ME,~\cite{Hespanha}) are widely adopted mathematical tools for analyzing and reconstructing the stochastic dynamics of the system~\cite{MunskyEtAl2009,ZechnerEtAl2012}.

\begin{figure}
\centering
\subfigure[\label{fig:ReporterScheme}]{\includegraphics[width=\columnwidth]{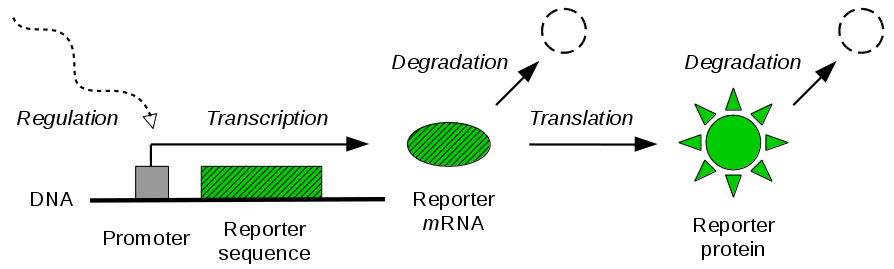}}\\
\subfigure[\label{fig:ReporterData}]{\includegraphics[width=\columnwidth]{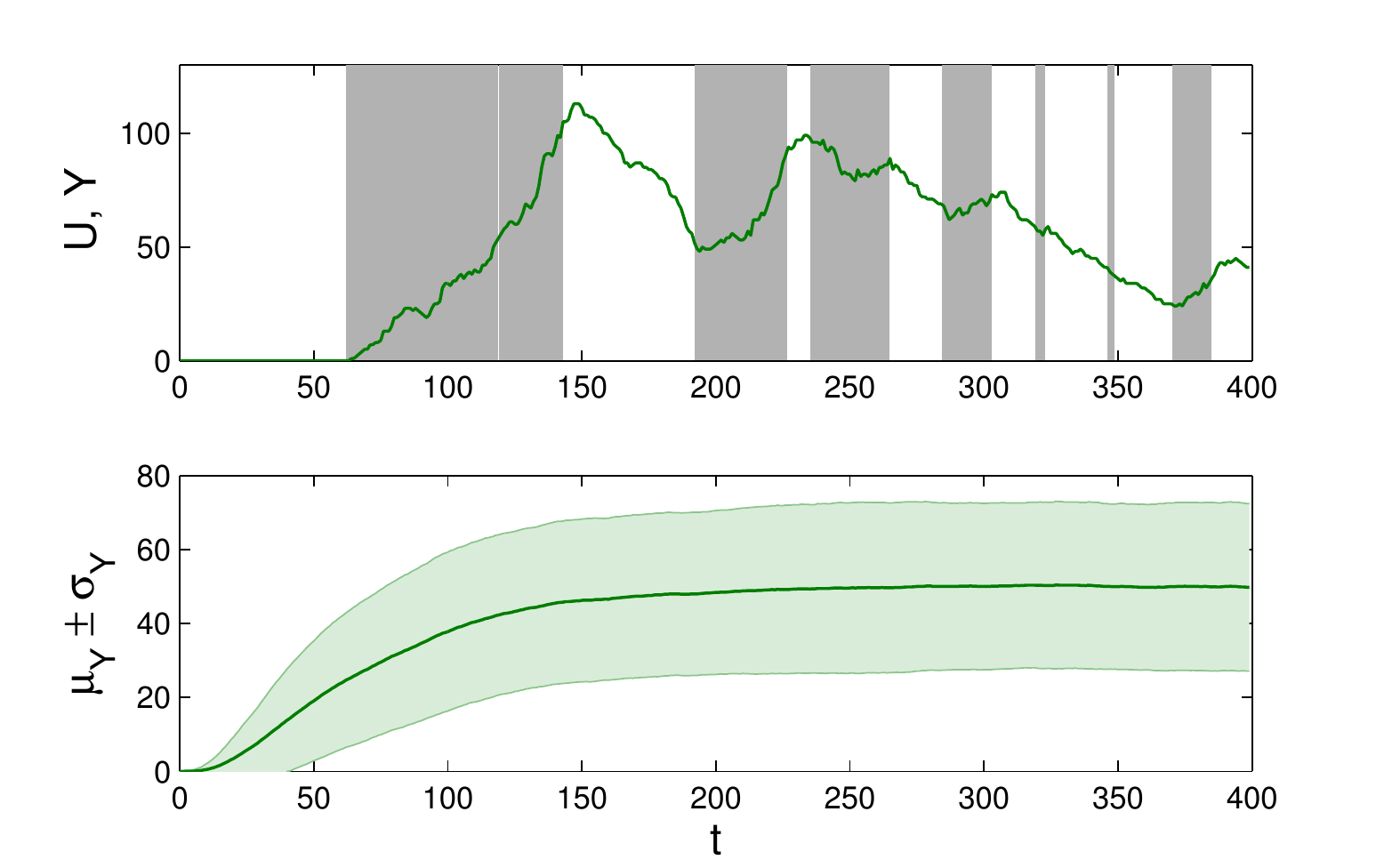}}
\caption{Reporter gene system. \subref{fig:ReporterScheme} Genetic construction and functioning of a reporter system. The coding sequence of a reporter protein is placed under the control of the promoter of a gene of interest. Upon gene expression, reporter \mRNA molecules are transcribed from the gene and further translated into visible (quantifiable) protein molecules. Both \mRNA and protein molecules are subject to degradation. \subref{fig:ReporterData} Simulated example of reporter system, assuming that no reporter molecules are present at time $0$. Top panel: Individual-cell profile of reporter abundance ($Y$, green line) in response to activation and deactivation over time of the promoter ($U$, resp. gray shades and white background) in the same cell; Bottom panel: Mean ($\mu_Y$, thick line) plus/minus standard deviation ($\sigma_Y$, thin lines and shaded region) of reporter abundance over a population of cells, each with its own promoter activation pattern, from population snapshot measurements. Time is in minutes, reporter abundance is in number of molecules.} \label{fig:grs}
\end{figure}

A widespread technique to monitor stochastic gene expression dynamics are reporter gene systems. These are genetic DNA constructs ensuring that new molecules of an easily quantifiable (for instance, fluorescent) protein are synthesized when a gene of interest is expressed. In a given cell, stochastic expression of the monitored gene leads to fluctuating trajectories of reporter abundance (refer to illustration in Fig.~\ref{fig:grs}), that can be traced in single cells by \emph{e.g.} time-lapse fluorescence microscopy~\cite{LlamosiEtAl2016,ZechnerEtAl2014}. Alternatively, population snapshot data obtained \emph{e.g.} by flow-cytometry~\cite{ZechnerEtAl2012,HasenauerEtAl2011} provide the distribution of reporter abundance levels within independent samples collected at different time instants from a population of cells. Because the data provide an indirect readout of the phenomenon of interest, a key challenge is to relate gene activation statistics with the reporter abundance data and, conversely, to infer the former from the latter.

From an engineering viewpoint, reporter systems can be seen as dynamical sensoring devices, with a random input (gene activation) driving stochastic dynamics that determine the sensor output (reporter abundance). Motivated by reporter systems, in this paper we address in more generality the analysis of stochastic reaction networks with input processes. We consider reaction rates that are affine in the state~\cite{Gadgil2005}. Under the assumption of stochastic causality~\cite{Granger,LindquistPicci,BowsherEtAl}, for an arbitrary input process with finite first- and second-order moments, 
we derive exact relationships between input and output statistics. When restricted to mean and variance, these equations constitute a generalization of the ME to the presence of stochastic inputs. We also derive equations relating the input and output autocovariance functions and, in stationary conditions, we provide a spectral characterization of the input-output transformation, showing analogies with and differences from linear filtering of stochastic processes~\cite{Gardner,LindquistPicci}. 

This first contribution relates with work on the analysis of noise propagation in biochemical networks.
Noise propagation in gene networks and dissection of different noise sources is treated in~\cite{BowsherEtAl,Swain2002,Pedraza2005,ThattaiVanOudenaarden2001,HilfingerEtAl2012}, among others. In~\cite{AustinEtAl2006}, spectral analysis is used to investigate the effects on gene expression noise of different gene regulatory configurations. In~\cite{WangEtAl2008}, spectral analysis based on the CME is explicitly performed for a specific parametric gene expression model. Effects of exogenous or unmodelled dynamics on the statistics of a reaction network are treated with various approaches in~\cite{HilfingerEtAl2016,LestasEtAl2010,YeungEtAl2013}. A Langevin approximation for the frequency-domain analysis of noise in genetic circuits is proposed in~\cite{SimpsonEtAl2003,Cox2006}. 
An approach to the analysis of stochastic reaction networks similar to ours is taken in~\cite{LestasEtAl2008}, but in absence of inputs and with a different focus. In a broader perspective, our analysis falls in the context of stochastic hybrid systems~\cite{Hespanha,LygerosPrandini}, providing results for a specific class of models that can be of interest, in particular, to the analysis, estimation and identification of CTMCs~\cite{NorrisBook}. 

Next, we exploit our general results to address reconstruction of input statistics from reporter data. 
Different from \emph{e.g.}~\cite{SuterEtAl,ZechnerEtAl2014,LlamosiEtAl2016}, where single-cell trajectories are presumed available, 
we focus on population snapshot data, which are experimentally easier to obtain.
When confined to population means, the problem reduces to deconvolution~\cite{DeNicolaoEtAl1997,PillonettoBell}, and has been addressed with success in a number of works~(see \emph{e.g.}~\cite{ZulkowerIMSB,FinkenstadtEtAl2008,Schelker2012}). In the stochastic setup, 
the problem is nontrivial, and has been addressed only indirectly. Most approaches are based on parameter estimation or model selection~\cite{KomorowskiEtAl2010,KomorowskiEtAl,HasenauerEtAl2011,NeuertEtAl2013}. Here we take a nonparametric approach, that is, we propose a method to reconstruct statistics of an arbitrary input process in absence of a parametric model governing its laws. We concentrate on the reconstruction of the autocovariance function of promoter activity, which is of particular interest since it conveys information about time scales and memory of the gene expression process. 
Nonparametric methods 
for population snapshot data are in their infancy~\cite{OconeEtAl,Hasenauer2014,RuessEtAl2011,CDC2015,HSB2015}. Different from correlation analysis~\cite{SodestromStoica}, time correlation of the output is not assumed available, which complicates the problem considerably. 
Yet nonparametric methods carry great potential, since they enable to decouple statistical characterization of gene expression from the mechanistic modelling of regulatory interactions. 

The paper, which is a vast extension and generalization of the preliminary work in~\cite{CinquemaniHSB2016}, is organized as follows.
The formal definition of stochastic reaction network is reviewed in Section~\ref{sec:modelling}. Moment equations for networks with stochastic inputs are derived in Section~\ref{sec:momentequations}. Based on this, spectral analysis is discussed in Section~\ref{sec:spectralanalysis}. In all these sections, the case study of reporter systems is further discussed as a running example. In Section~\ref{sec:reconstruction}, based on the previous results, we develop a method for the nonparametric estimation of a stationary gene promoter autocovariance function from transient population snapshot (mean and variance) reporter gene data. Theoretical results as well as the performance of the reconstruction method are demonstrated via numerical simulations in Section~\ref{sec:example}. Section~\ref{sec:conclusions} concludes the paper with a final discussion and perspectives of the work.
All mathematical proofs are reported in Appendix~\ref{asec:proofs}.
\par
\emph{Notation:} $\NN$, $\ZZ$, $\RR$, $\RR_{\geq 0}$ and $\CC$ denote natural, integer, real, nonnegative real and complex numbers, respectively. For a set $T\subset\RR$, $\ind_T(\cdot)$ is the indicator function of $T$, and $\ind(\cdot)$ is the unit step function $\ind_{[0,+\infty)}(\cdot)$. For three random vectors $X$, $Y$ and $F$, $\EE[X|F]$ denotes conditional expectation of $X$ given $F$, $\Cov(X,Y|F)=\EE[(X-\EE[X|F])(Y-\EE[Y|F])^T|F]$ (superscript ``$^T$'' denoting transposition) and $\Var(X|F)=\Cov(X,X|F)$. $\Cov(X,Y)$ and $\Var(X)$ are defined similarly, with conditional expectations $\EE[\,\cdot\,|F]$ replaced by simple expectations $\EE[\,\cdot\,]$. $\PP[\,\cdot\,]$ denotes probability of an event.

\section{Stochastic reaction networks with inputs}

\label{sec:modelling}

A reaction network is a family of chemical species and reactions that occur among them in a given reaction volume. Consider a network with $n$ species $\S_1,\ldots, \S_n$ and $m$ reactions $\R_1,\ldots, \R_m$. To our purposes, the reaction network is fully described by a stoichiometry matrix $S\in\ZZ^{n\times m}$ and by a vector of reaction rates $w\in\RR^{m}$. The $i$th row, $j$th column entry of $S$ denotes the net change in the number of molecules of $\S_i$  when reaction $\R_j$ takes place. 
Under suitable assumptions on the reaction volume, reaction rates depend on the abundance of the different species as dictated by the laws of mass action~\cite{Gillespie1992}, and describe the propensity (limiting probability over an infinitesimal time period) by which the different reactions take place. Correspondingly, vector $$X(t)=\begin{bmatrix} X_1(t) & \cdots & X_n(t)\end{bmatrix}^T,$$ where $X_i(t)$ is the number of molecules of $\S_i$ at time $t$, describes the system state at time $t$ and follows the laws of a CTMC. 

We consider reaction networks with rates of the form
\begin{equation}
\label{eq:rates}
w^f(t)=W X(t)+f(t),
\end{equation}
with $W\in\RR_{\geq 0}^{m\times n}$, where $f:\RR\to \RR_{\geq 0}^{m}$ is a piecewise continuous function. This form is peculiar of reaction networks comprising zero- or first-order reactions, and the starting point for the approximate description of more complex reaction dynamics (see \textit{e.g.}~\cite{ThattaiVanOudenaarden2001}).  
Possible generalizations of this assumption will be considered in the discussion of Section~\ref{sec:conclusions}.

We are interested in the general case where $f$ is the random outcome of a stochastic process $F$ that is a causal input of the system. That is, we assume absence of feedback from $X$ to $F$, so that, independent of the specific outcome of $F$, reaction rates~\eqref{eq:rates} can be written as
\begin{equation}
\label{eq:stochrates}
w(t)=WX(t)+F(t).
\end{equation}
We assume that the first- and second-order moments of $F$ are uniformly bounded. Note that this includes the case where some (or even all of the) components of $F(t)$ are deterministic. In agreement with the nonnegativity of the elements of $f(t)$, we assume that $\EE[F(t)]\geq 0$ elementwise for all $t$. 

\subsection{Case study: Reporter gene systems}
\label{sec:reportermodelling}

Refer to Fig.~\ref{fig:ReporterScheme}. Gene expression kinetics can be described
by the reaction system
\begin{equation}
\label{eq:randtel}
\begin{aligned}
 \R_1:~&\emptyset \xrightarrow{k_M\cdot U} M  &\R_2:~&M\xrightarrow{d_M}\emptyset \\
 \R_3:~&M\xrightarrow{k_P} M+P  &\R_4:~&P\xrightarrow{d_P}\emptyset 
\end{aligned}
\end{equation}
\cite{Friedman2006,KaernEtAl2005} where $M$ and $P$ denote \mRNA and protein species, respectively. Reaction $\R_1$ represents transcription of the coding sequence of the gene into one \mRNA molecule, while reaction $\R_3$ represents translation of one \mRNA into one new molecule of the protein $P$ coded by the gene. Reactions $\R_2$ and $\R_4$ describe the degradation of the \mRNA and protein molecules, respectively. 

In the context of this paper, $P$ is the fluorescent reporter protein. 
We will not distinguish between immature (invisible) and mature (visible) protein molecules. If necessary (e.g. for slow, stochastic maturation), an additional first-order reaction $P\to P_{mature}$ can be included in the model (along with $P_{mature}\to\emptyset$) to account for protein maturation (and mature protein degradation). Without loss of generality, we ignore possible constant factors converting molecule abundance into observed fluorescence level.

In individual cells, reactions are best described as random events, so that stochastic network modelling applies.
In accordance with the standard random telegraph model~\cite{Paulsson2005157}, $U$ is a binary process such that, at time $t$, $U(t)=1$ if the gene is active, while $U(t)=0$ if the gene is inactive. Propensities of reactions $\R_1$--$\R_4$ are determined by the rate parameters $\theta=(k_M,d_M,k_P,d_P)$, which we assume to be positive constants. In this model, transcription (reaction $\R_1$) occurs at a rate $k_m$ only when the gene is active, while it does not occur when the gene is inactive. Note however that the expression $k_M\cdot U$ for the rate of $\R_1$ may accommodate  more complicated scenarios, such as the existence of multiple on-states~\cite{NeuertEtAl2013}, and admits a much larger interpretation where $U$ is any form of extrinsic noise~\cite{KomorowskiEtAl2010,Swain2002}. Most results in later sections are developed in such full generality. 
 
Let $X_1(t)$ and $X_2(t)$ denote the abundance of $M$ and of $P$ at time $t$, in the same order.
Then, from~\eqref{eq:randtel}, 
$$
S=\begin{bmatrix}
    1 & -1 & 0 & 0  \\ 
     0 & 0 & 1 & -1 \\
\end{bmatrix}
$$
and, from the laws of mass-action~\cite{Gillespie1992},
$$w(t)=
\begin{bmatrix}
k_M U(t) & d_M X_1(t) & k_P X_1(t) & d_P X_2(t)
\end{bmatrix}^T,
$$
which is in the form~\eqref{eq:stochrates} provided the definitions
\begin{align}
\label{eq:reporterF}
W&=\begin{bmatrix}
    0 & 0 \\
    d_M & 0 \\
    k_P & 0 \\
    0 & d_P
\end{bmatrix}, &F(t)&=
\begin{bmatrix}k_M U(t) \\ 0 \\ 0 \\ 0
\end{bmatrix}.
\end{align}
In this case, from a biological standpoint~\cite{BowsherEtAl}, absence of feedback from $X$ to $F$ is supported by the fact that fluorescence reporter proteins are by choice not part of the native proteome of the organism under study, and thus not part of specific gene expression regulatory mechanisms. In addition, from the viewpoint of experimental design, it is a prominent effort of synthetic biology to minimize cross-talking of the engineered biochemical modules (in this case, reporter systems) with the native cellular dynamics. 

From an engineering perspective, a gene reporter system can be seen as a sensoring device that transforms the process of interest $U$ into a measurement process $Y$ via a stochastic dynamical transformation. Thus, following standard conventions, we define $Y=X_2$ to emphasize the role of the reporter protein process as the output $Y$ of a system with input $U$ and state $X$. 

\section{Generalized Moment Equations}
\label{sec:momentequations}

For the reaction networks of Section~\ref{sec:modelling} with a random input $F$, we seek equations for the mean, variance, and autocovariance functions of $X$. We will prove that these equations can be written in closed form in terms of analogous statistics of $F$.
To achieve this, we will first consider the conditional statistics of $X$ given a generic outcome of $F$ and initial condition $X(0)$, and then proceed by marginalization. 

For a given profile $F=f$ and initial condition $X(0)=x_0$, define the conditional mean, covariance matrix and autocovariance (matrix) function 
\begin{align*}
\mu^{f,x_0}(t)&=\EE[X(t)|f,x_0], \\
\Sigma^{f,x_0}(t)&=\Var\big(X(t)|f,x_0\big), \\
\rho^{f,x_0}(\newt,t)&=\Cov\big(X(\newt),X(t)|f,x_0\big),
\end{align*}
in the same order
(since $\rho(\newt,t)=\rho(t,\newt)^T$ by its very definition, we can restrict attention to $\newt\geq t$). For the conditioned process, reaction rates obey~\eqref{eq:rates} by the causality assumption. Hence, differential equations for the evolution of $\mu^{f,x_0}$ and $\Sigma^{f,x_0}$ are provided by the well-known ME for state-affine rates, which are extensively utilized in the literature (see \textit{e.g.}~\cite{Hespanha,MunskyEtAl2009,RuessEtAl2011}).  
Partial differential equations for $\rho^{f,x_0}(\newt,t)$ for the case of affine rates can instead be found in~\cite{LestasEtAl2008} (see also~\cite{WangEtAl2008} for a specific case study). To discuss these equations, let us introduce some notation. Define 
\begin{equation}
\label{eq:kernel}
\ell(t)=\exp(SWt) \ind(t)
\end{equation} 
and, for any matrix function $g(\cdot)$ such that the integrals are well-defined, the linear operations
\begin{align*}
\Lscr_t g&=\int_0^{+\infty} d\tau~\ell(t-\tau) g(\tau),\\
\Lscr_t^* g&=\int_0^{+\infty} d\tau~g(\tau)\ell(t-\tau)^T,\\
\Lscr^\circ_t g&=\int_0^{+\infty} d\tau~\ell(t-\tau) g(\tau) \ell(t-\tau)^T.\\
\end{align*}
For a matrix function $g(\cdot,\cdot)$ depending on two arguments, we let $\Lscr_t$ operate on the first argument and $\Lscr_t^*$ on the second argument, such that, for suitable $g'$ and $g''$, $\Lscr_t g(\cdot,\newt)=g'(t,\newt)$ and $\Lscr^*_t g(\newt,\cdot)=g''(\newt,t)$.
We also formally define the linear operators $\Lscr$, $\Lscr^*$ and $\Lscr^\circ$ transforming $g$ into functions over $\RR$ defined by $(\Lscr g) (t)=\Lscr_t g$, $(\Lscr^* g)(t)=\Lscr^*_t g$ and $(\Lscr^\circ g)(t)=\Lscr^\circ_t g$.

The following result recapitulates the differential equations for mean, variance and autocovariance in the special case of a fixed initial state $x_0$, and reports their integral solution.
\begin{proposition}
\label{thm:conditionalmoments}
For any $t$ and $\newt\geq t$, it holds that
\begin{align*}
\dot \mu^{f,x_0}(t)& =SW\mu^{f,x_0}(t)+S  f(t),\\
\dot \Sigma^{f,x_0}(t)& = SW\Sigma^{f,x_0}(t)+ \Sigma^{f,x_0}(t)W^TS^T + Q^{f,x_0}(t),\\
\frac{\partial}{\partial\newt}\rho^{f,x_0}(\newt,t)&=SW\rho^{f,x_0}(\newt,t),
\end{align*}
with $\mu^{f,x_0}(0)=x_0$, $\Sigma^{f,x_0}(0)=0$ and
$\rho^{f,x_0}(t,t)=\Sigma^{f,x_0}(t)$, where $Q^{f,x_0}(t)=S\emph{\diag}\big(W\mu^{f,x_0}(t)+ f(t)\big)S^T$.
The solutions are
\begin{align}
\label{eq:condmean}
\mu^{f,x_0}(t)&=\ell(t)  x_0+\Lscr_t(S   f), \\
\label{eq:condvar}
\Sigma^{f,x_0}(t)&=\Lscr^\circ_t(Q^{f,x_0}), \\ 
\label{eq:condautocov}
\rho^{f,x_0}(\newt,t)&= \ell(\newt-t)\Sigma^{f,x_0}(t).
\end{align}
\end{proposition} 
Next define the (non-conditional) moments
\begin{align*}
\mu_F(t)&=\EE[F(t)], &\rho_F(\newt,t)&=\Cov\big(F(\newt),F(t)\big), \\
\mu(t)&=\EE[X(t)], &\rho(\newt,t)=&\Cov\big(X(\newt),X(t)\big) \\
\end{align*}
as well as $\Sigma(t)=\rho(t,t)$, and denote $\mu_0=\mu(0)$, $\Sigma_0=\Sigma(0)$. 
(Note that to simplify notation, throughout the manuscript, the statistics of process $X$, \emph{e.g.} $\mu$, are written without a subscript ``$X$'', contrary to analogous statistics of other processes, \emph{e.g.} mean $\mu_F$ of process $F$.)
Also define
$\xi_F(t)=\Cov\big(X(0),F(t)\big)$. We assume that $\mu_0$ and $\Sigma_0$ are well-defined (finite). Together with the assumptions on $F$, this also implies that $\xi_F$ is uniformly bounded. Marginalization of the conditional moments~\eqref{eq:condmean}--\eqref{eq:condautocov} eventually leads to the following result.
\begin{proposition}
\label{thm:moments}
For any $t\geq 0$ and $\newt\geq t$ it holds that
\begin{align*}
\mu(t)&=\ell(t)\mu_0+\Lscr_t(S   \mu_F), \\
\rho(\newt,t)
&=\ell(\newt)\Sigma_0\ell(t)^T+\Lscr^*_t\Lscr_\newt(S   \rho_F  S^T)+\\
&\qquad\ell(\newt) \Lscr^*_t(\xi_F  S^T)+ \Lscr_\newt (S  \xi_F^T)  \ell(t)^T+\\
&\qquad\ell(\newt-t)  \Lscr^\circ_t\left(S\emph{\diag}\big(W  \mu+\mu_F\big)S^T\right).
\end{align*}
\end{proposition}
It can be appreciated that $\mu$ and $\rho$ are the result of linear transformations of the mean and autocovariance of $F$, that is $\mu_F$ and $\rho_F$, plus terms associated with the initial distribution of $X(0)$. For $X(0)=0$, in particular,
the expressions are in close analogy with the transformations that would be operated on $\mu_F$ and $\rho_F$ by a linear filter with convolution kernel $\ell(\cdot)S$ applied to $F$, with the exception of the term in $\rho$ depending on $\Lscr^\circ_t$.
We will come back on this point in Section~\ref{sec:spectralanalysis}.
\begin{corollary}
\label{thm:diffform}
For any $t\geq 0$ and $\newt\geq t$, and any $\mu_0$ and positive semi-definite $\Sigma_0$, it holds that
\begin{align}
\label{eq:gendiffeqmu}
\dot\mu(t)&=SW  \mu(t)+S   \mu_F(t), \\
\label{eq:gendiffeqSigma}
\dot\Sigma(t)&=SW  \Sigma(t)+\Sigma(t)  W^TS^T +Q(t) + \\
&\qquad V_\xi(t,t)+V_\xi^T(t,t)+V_\rho(t,t)+V_\rho^T(t,t), \nonumber \\
\label{eq:gendiffeqXi}
\frac{\partial}{\partial\newt} \rho(\newt,t)&=SW  \rho(\newt,t)+V_\xi(\newt,t)+V_\rho(\newt,t),
\end{align}
with $\mu(0)=\mu_0$, $\Sigma(0)=\Sigma_0$ and $\rho(t,t)=\Sigma(t)$, where
\begin{align*}
Q(t)&=S  \emph{diag}\big(W\mu(t)+\mu_F(t)\big)  S^T, \\
V_\xi(\newt,t)&=S  \xi_F(\newt)^T\ell(t)^T, \\
V_\rho(\newt,t)&=\Lscr^*_t(S  \rho_F(\newt,\cdot)  S^T).
\end{align*}
\end{corollary}
This result follows straight from Proposition~\ref{thm:moments} by taking derivatives. By comparison with Proposition~\ref{thm:conditionalmoments}, it shows that the evolution of first- and second-order moments resembles that of the conditioned process moments, except for $\mu_F(t)$ in place of $f(t)$ and for the input terms $V_\xi$ and $V_\rho$. These input terms characterize the additional variability of $X$ as a result of the variability of $F$.

\subsection{GME for reporter systems}
\label{sec:momentsrandomregulator}

For the reporter model of Section~\ref{sec:reportermodelling}, 
it is immediately found that
$$SW=\begin{bmatrix}-d_M & 0 \\ k_P & -d_P\end{bmatrix}.$$ 
The explicit form of the matrix exponential~\eqref{eq:kernel} is given by standard formulas. Two different expressions are obtained depending on whether $d_M=d_P$ or $d_M\neq d_P$. Biologically speaking, the most relevant case is $d_P<d_M$, since \mRNA molecules are usually less stable than proteins. We will not make this assumption for the time being, but come back to it for certain explicit calculations later on. In view of the specific form of $F$ in Eq.~\eqref{eq:reporterF}, it is of interest to detail the formulas in terms of $U$.
Define 
\begin{align*}
\mu_U(t)&=\EE[U(t)], \\ 
\rho_U(\newt,t)&=\Cov\big(U(\newt),U(t)\big), \\
\xi_U(t)&=\Cov\big(X(0),U(t)\big).
\end{align*}
By straightforward calculations, one finds that
\begin{equation}
\label{eq:Q}
Q(t)=\begin{bmatrix} k_M \mu_U(t)+d_M\mu_1(t) & 0 \\ 0 & k_P\mu_1(t)+d_P\mu_2(t)\end{bmatrix},
\end{equation}
with $\mu_i$ the $i$th element of $\mu$, 
as well as the expressions of $V_\xi(\newt,t)$ and $V_\rho(\newt,t)$, respectively given by
\begin{equation}
\label{eq:V}
\begin{bmatrix}k_M\xi_U(\newt)^T\ell(t)^T \\ 0_{1\times 2}\end{bmatrix}, \quad
\begin{bmatrix}k_M^2\Lscr^*_t([\,\rho_U(\newt,\cdot)~0\,]) \\ 0_{1\times 2}\end{bmatrix}.
\end{equation}
Together with $\mu_F(t)=[\,k_M \mu_U(t)~0~0~0\,]^T$, these expressions can be used to specialize the results in Proposition~\ref{thm:moments} and Corollary~\ref{thm:diffform}. In particular, it is possible to rewrite the differential equations~\eqref{eq:gendiffeqmu}--\eqref{eq:gendiffeqSigma} in the vectorized form often encountered in the literature~\cite{CDC2015,MunskyEtAl2009}, that is, a system of linear equations with state vector formed by the non-redundant entries of $\mu$ and $\Sigma$. Different from this literature, the resulting vector field would also include terms from $V_\rho$ and $V_\xi$, accounting for the stochastic nature of $U$. We will not detail this vectorized form. The above expressions will instead be utilized for the developments of Section~\ref{sec:spectrumrandomregulator} and~\ref{sec:reconstruction}.

\section{Spectral analysis of networks with inputs}
\label{sec:spectralanalysis}

Starting from the general results of Section~\ref{sec:momentequations}, suppose now $F$ is second-order stationary~\cite{Gardner,LindquistPicci}, that is, $\mu_F(t)$ is constant and $\rho_F(\newt,t)$ depends only on the difference $\newt-t$. Irrespective of $t$ we may thus define
$\bar \mu_F=\mu_F(t)$, $\bar\rho_F(\delta)=\rho_F(t+\delta,t)$. For two generic functions $g$ and $g'$ over $\RR$ taking matrix values of compatible size, let $g\conv g'$ denote convolution, $(g\conv g')(t)=\int_{-\infty}^{+\infty}d\tau~g(t-\tau)g'(\tau)$. 
\begin{proposition}
\label{thm:stationarymoments}
Assume that the square matrix $SW$ is strictly stable, that is, all its eigenvalues $\lambda$ obey $\text{Re}(\lambda)<0$.
Then, for $t\to+\infty$ and any $\delta\geq 0$, $\mu(t)\to\bar \mu$ and $\rho(t+\delta,t)\to \bar \rho(\delta)$, where, for $\ell_-(\cdot)=\ell(-\cdot)$,
\begin{align*}
\bar \mu&=-(SW)^{-1}S \bar\mu_F, \\ 
\bar \rho(\delta)
&=\big[\ell\conv(S   \bar \rho_F  S^T)\conv\ell_-^T\big](\delta)+\ell(\delta)  \bar \Sigma^\circ, 
\end{align*} 
where $\bar \Sigma^\circ=\Lscr^\circ_{\infty}(\bar Q)$ is the unique symmetric
solution of
\begin{align}
\label{eq:sigmacirc}
0&=SW \bar \Sigma^\circ+\bar \Sigma^\circ W^TS^T+\bar Q, 
\end{align}
with $\bar Q=S\emph{\diag}\big((I-W(SW)^{-1}S) \bar\mu_F \big)S^T$.
\end{proposition}
Thus, a second-order stationary process $F$ leads to a second-order stationary process $X$ at least asymptotically.
The expression of $\bar \rho(\delta)$ for $\delta<0$ follows from the equality $\rho(t+\delta,t)=\rho(t,t+\delta)^T$, such that $\bar \rho(\delta)=\bar \rho(-\delta)^T$. For $\delta=0$, the stationary covariance matrix $\bar\Sigma = \bar \rho(0)$ is obtained. 

The expression of the stationary autocovariance function $\bar \rho$ obtained in Proposition~\ref{thm:stationarymoments} is composed of a convolutional term, which is typical of linear filtering, and of an additional term given by the exponential factor $\ell(\cdot)$ times a constant matrix.
Starting from this expression, one may quantify in the frequency domain how the (matrix) power spectral density of $F$ is reflected into that of $X$. Recall that the power spectral density of a (second-order) stationary process with absolutely integrable (matrix) autocovariance function $r(\cdot)$ is given by the Fourier transform $\F r$~\cite{Gardner,LindquistPicci}, 
$$
(\F r)(i\omega)=\int_{-\infty}^{+\infty}d\delta\, e^{-i\omega\delta}r(\delta),
$$
with $\omega\in\RR$ and $i$ the imaginary unit.
For $\bar\rho_F$ absolutely integrable, let $R_F=\F \bar\rho_F$ be the spectral density of $F$. By well-known properties of Fourier analysis, in view of the expression of $\bar \rho$, the spectral density $R=\F\bar \rho$ of $X$ is given by
\begin{multline}
\label{eq:spectrum}
R(i\omega)=L(i\omega) S   R_F(i\omega)  S^T L(-i\omega)^T+\\ L(i\omega)  \bar \Sigma^\circ + \bar \Sigma^\circ  L(-i\omega)^T,
\end{multline}
where $L=\F\ell$. By the definition of $\ell$ in~\eqref{eq:kernel} and the assumption that $SW$ is strictly stable, the entries of $L(i\omega)$ are finite for all $\omega\in\RR$. Therefore $R$ is also well-defined.
For $s\in\CC$, let us interpret $L(\sl)S$ as the transfer function of a linear filter with impulse response $\ell(\cdot)S$. Then the first term of~\eqref{eq:spectrum} is the input-output transformation of the power spectrum operated by this linear filter, 
while the second term introduces an unusual component in the output spectrum associated with the mean of $F$ via $\bar\Sigma^\circ$.
In other words, in terms of first- and second-order statistics, we may interpret stochastic reaction networks as the superposition of a linear filter operating on the input process $F$, plus a stochastic component introduced by the randomness of the network reactions. This interpretation is further elaborated below.

\subsection{Spectral analysis of reporter gene systems}
\label{sec:spectrumrandomregulator}

\begin{figure}
\centering
\includegraphics[width=\columnwidth]{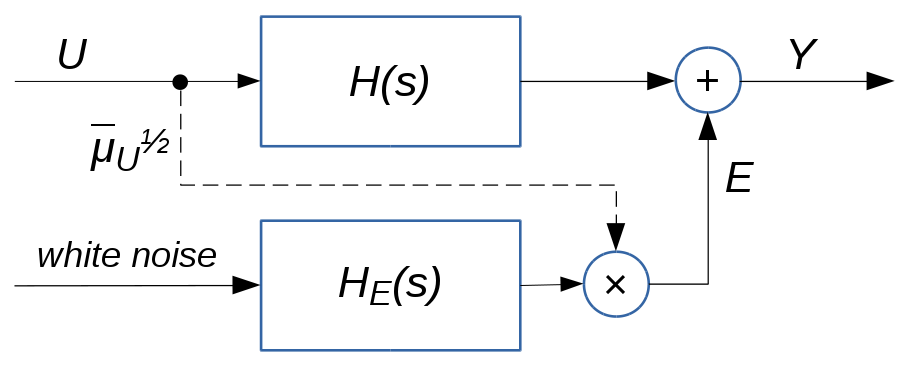}
\caption{Filter representation of the reporter system. The dashed line branching from $U$ indicates a statistical property of the process.}\label{fig:reporterfilter}
\end{figure}
For the reporter gene model introduced in Section~\ref{sec:reportermodelling}, we further develop the results of Section~\ref{sec:momentsrandomregulator} under the assumption that $SW$ is strictly stable. This holds whenever the rate constants $d_M$ and $d_P$ are both strictly positive, which is generally the case.
In view of Proposition~\ref{thm:stationarymoments}, for $U$ stationary, we may then consider the stationary statistics of $X$.
While the results of Proposition~\ref{thm:stationarymoments} could be spelled out in full detail for the case at hand, we are mostly interested in the spectrum of the output (reporter protein) process $Y$ in terms of the spectrum of the input process $U$. 
Let $\bar \mu_U$ and $\bar\rho_U(\cdot)$ be the stationary mean and autocovariance function of $U$, in the same order.
Also let $R_U$ and $R_Y$ respectively denote the Fourier transforms of $\bar\rho_U$ and of the stationary autocovariance function of $Y$.

\begin{proposition}
\label{thm:reporterspectrum}
Let $U$ and $X$ be (second-order) stationary. The power spectral density of $Y$ is given by
\begin{equation}
R_{Y}(i\omega)=H(i\omega)H(-i\omega)R_U(i\omega)+R_E(i\omega)
\end{equation}
where, for $\alpha=k_P/(d_M+d_P)$, $r_P=k_P/d_P$ and $r_M=k_M/d_M$,
\begin{align*}
H(i\omega)&=\frac{k_Pk_M}{(d_P+i\omega)(d_M+i\omega)}, \\
R_E(i\omega)&=r_M \left[\frac{\alpha k_P}{(d_P+i\omega)(d_M+i\omega)}+\frac{r_P(1+\alpha)}{(d_P+\sl)}+\right.\\
&\qquad \left.\frac{\alpha k_P}{(d_P-i\omega)(d_M-i\omega)}+\frac{r_P(1+\alpha)}{(d_P-i\omega)}\right]\bar \mu_U.
\end{align*}
\end{proposition}

Notice that, from the definition of $U$ as a binary process and in agreement with the assumptions on $F$, $\bar \mu_U\geq 0$. It can thus be appreciated that the reporter system acts on the spectrum of $U$ as a linear dynamical system with transfer function $H(\sl)$, except for the additional term $R_E(i\omega)$, proportional to the input process mean $\bar \mu_U$, that reflects the stochasticity of the reporter system. From a signal processing viewpoint, we may thus interpret gene reporters as linear filters introducing colored measurement error $E$ with mean zero and power spectrum $R_E(i\omega)=H_E(i\omega)H_E(-i\omega)\bar\mu_U$, for a suitable spectral factor $H_E(i\omega)$. A pictorial view of this interpretation is given in Fig.~\ref{fig:reporterfilter}.

\section{Reconstruction of promoter statistics from population snapshot data}
\label{sec:reconstruction}

Reporter systems provide an indirect readout of gene activation and deactivation. Reconstruction of promoter activity statistics is thus the core problem in reporter gene data analysis. The relationships between second-order moments of input and state (output) processes that were established in the previous sections enable one to address a variety of reconstruction problems. The linearity of these relationships makes mathematical treatment very neat. 
In this section, we concentrate on the reconstruction of the statistics of a promoter process from population snapshot data. To do this, we will elaborate on the results of Section~\ref{sec:momentsrandomregulator}. 
We assume that the reporter parameters $\theta$ are known, with $d_P < d_M$, and consider a scenario where $U$ is in (or has reached) stationary conditions.
For the reasons exposed in the introduction, we are especially interested in the (nonparametric) reconstruction of the (stationary) autocovariance function $\bar\rho_U$. In the light of~\eqref{eq:gendiffeqSigma}, with $V_\rho$ as in~\eqref{eq:V}, $\bar\rho_U$ acts as a forcing input on the dynamics of $\Sigma$. This prompts us to address estimation of $\bar\rho_U$ from measurements of the transient of $\Sigma$. To avoid that $\Sigma$ is in steady-state, one must ensure (mathematically and experimentally) that $X$ is not stationary. Here we develop a method for the case where $X(0)=0$, which is especially convenient since it avoids stationarity of $X$ and implies that $V_\xi=0$ in~\eqref{eq:gendiffeqSigma}. 

We assume that snapshot measurements of the population mean, $\tilde \mu_Y$, and of the variance, $\tilde \sigma^2_Y$, are available at increasing measurement times $t_k\geq 0$, with $k=0,\ldots, M-1$. 
Because these empirical statistics are drawn from large and independent samples (typically in flow-cytometry, $10^4$ or more cells per sample), thanks to the law of large numbers, measurements can be modelled as
\begin{align*}
\tilde\mu_Y(t_k)&=\mu_2^*(t_k)+e^\mu_k,&\tilde\sigma_Y^2(t_k)=&\Sigma_{2,2}^*(t_k)+e^\sigma_k,
\end{align*} 
where $\mu_2^*$ and $\Sigma_{2,2}^*$ denote the true mean and variance of process $Y$, and the errors $(e^\mu_k,e^\sigma_k)$ are approximately Gaussian, zero-mean and uncorrelated across $k$~\cite{ZechnerEtAl2012}. Subscript indices emphasize that $\mu_2^*$ and $\Sigma_{2,2}^*$ are entries of the true mean vector and covariance matrix of $X$, denoted with $\mu^*$ and $\Sigma^*$. (In the present context, superscript ``\,*\,'' is used to clearly distinguish true process statistics from candidate solutions).

The moment equations of the system reduce to 
\begin{align}
\label{eq:muFrom0}
\dot\mu(t)&=SW\mu(t)+\begin{bmatrix} k_M\bar \mu_U & 0\end{bmatrix}^T, \\
\label{eq:SigmaFrom0}
\dot \Sigma(t)&=SW \Sigma(t)+\Sigma(t) W^TS^T +Q(t) + V_\rho(t)+V_\rho^T(t),
\end{align}
with $\mu(0)=0$ and $\Sigma(0)=0$, where $V_\rho(t)$ is shorthand for $V_\rho(t,t)$. 
Using the fact that $d_M\neq d_P$, the expression of $V_\rho(t)$ in~\eqref{eq:V} becomes
\begin{equation}
\label{eq:Vstationary}
\begin{bmatrix} k_M^2 \I_r(t;d_M)& \frac{k_M^2 k_P}{d_P-d_M}\big(\I_r(t;d_M)-\I_r(t;d_P)\big)\\ 0 & 0\end{bmatrix}
\end{equation}
with $r=\bar\rho_U$, where, for a generic measurable function $r:\RR\to\RR$ and any constant $d_\circ>0$,
$\I_r(t;d_\circ)=\int_0^t d\delta \,r(\delta)\exp(-d_\circ\cdot\delta)$. 

We assume that the model is exact, that is, $\mu^*$ and $\Sigma^*$ are the solution of~\eqref{eq:muFrom0}--\eqref{eq:SigmaFrom0} corresponding to some true mean $\bar\mu_U^*$ and autocovariance function $\bar\rho_U^*$, and denote with $Q^*$ and $V_\rho^*$ the corresponding instances of $Q$ and $V_\rho$.
From~\eqref{eq:Q}, $Q(t)$ depends on $\bar\mu_U$ and $\mu(\cdot)$. In turn, $\mu(\cdot)$ is determined by $\bar\mu_U$ via~\eqref{eq:muFrom0}, so that $Q(t)$ is essentially a function of the constant $\bar\mu_U$. The true value of the latter is especially easy to reconstruct from data, since it suffices to fit the observations $\tilde \mu_Y(\cdot)$ with the solution of~\eqref{eq:muFrom0} as a function of the scalar constant $\bar \mu_U$. We will not discuss this trivial problem further, and simply assume that $Q^*(\cdot)$ is known.

Let $\Sigma(t|r,Q)$ be the solution of~\eqref{eq:SigmaFrom0}--\eqref{eq:Vstationary} for some generic $Q$ and $r$.
Define $v(t|r)=\Sigma_{2,2}(t|r,0)$
and $v_0(t)=\Sigma_{2,2}(t|0,Q^*)$. 
In view of the linearity of~\eqref{eq:SigmaFrom0} in $V_\rho$ and $Q$, and the linearity of~\eqref{eq:Vstationary} in $r$, 
$v(\cdot|r)$ is a linear functional of $r$, and it holds that $\Sigma_{2,2}^*=v(t|\bar \rho_U^*)+v_0(t)$.
Thus, in principle,
estimation of $\bar \rho_U$ may be formulated  
as the linear least-squares problem
\begin{equation}
\label{eq:generalreconstruction}
\inf_{r\in \C}\sum_{k=0}^{M-1}\alpha_k^2\cdot\left(\tilde \sigma^2_Y(t_k)-v_0(t_k)-v(t_k|r)\right)^2
\end{equation}
where $\C$ is the convex cone of stationary autocovariance (equivalently, positive-semidefinite) functions over $\RR$, and, for every $k$,
weight $\alpha_k^2$ is fixed to the inverse of the variance of $e^\sigma_k$, which can itself be estimated from the data as in~\cite{ZechnerEtAl2012}.
However, the problem is ill-posed due to the sampling of the data and the infinite-dimensional nature of $\C$~\cite{Bertero,DeNicolaoEtAl1997}. We thus recast the problem into regularized estimation using a finite-dimensional approximation of $\C$. 

Let $\{r_1,\ldots,r_N\}$ be a fixed set of symmetric measurable functions. Due to the linearity of $v$, 
\begin{align*}
v(t|c_1r_1+\ldots+c_Nr_N)&=\sum_{\ell=1}^N c_\ell v_\ell(t),&v_\ell(t)&=v(t|r_\ell).
\end{align*}
Defining $\R(\cdot)=[\,r_1(\cdot)~\cdots~r_N(\cdot)\,]$ and $\V(\cdot)=[\,v_1(\cdot)~\cdots~v_N(\cdot)\,]$, we may then consider 
estimates $\hat\rho_U$ of $\bar\rho_U$ of the form $\R(\cdot)\hat c$, where $\hat c$ is a solution of
\begin{align}
\label{eq:finitereconstructionobjective}
\inf_{c\in\RR^N}~&\sum_{k=0}^{M-1}\alpha_k^2\cdot\left(\tilde \sigma^2_Y(t_k)-v_0(t_k)-\V(t_k) c\right)^2+\gamma\Q(c) \\
\label{eq:finitereconstructionconstraints}
\text{s.t.}~&\R(\cdot)c\in\C,
\end{align}
where $\Q(c)$ is a quadratic penalty function promoting regular solutions, and $\gamma\geq 0$~\cite{DeNicolaoEtAl1997}.
Optimization~\eqref{eq:finitereconstructionobjective}--\eqref{eq:finitereconstructionconstraints} is a linear least-squares problem with convex constraints, which can be solved efficiently~\cite{BoydVandenberghe}. Of course,
the choice of the $r_\ell$ and of $\gamma\Q$ will determine accuracy and regularity of the estimate $\hat\rho_U(\cdot)$.
From an implementation viewpoint, a convenient choice of the $r_\ell$ may simplify the computation of the $v_\ell$. In particular, one may ensure that the integrals $\I_{r_\ell}(\,\cdot\,; d_\circ)$ appearing in $\eqref{eq:Vstationary}$ can be calculated explicitly. In addition, depending on the choice of the $r_\ell$, a more explicit form for the contraints~\eqref{eq:finitereconstructionconstraints} should be determined.
One convenient choice is discussed in detail in the next section.

\subsection{Solution by expansion over indicator functions}

First of all notice that, for any $r$, the quantities $v(t_0|r),\ldots,v(t_{M-1}|r)$ depend on $r(\delta)$ only for $\delta\in T=[0,t_{M-1})$. 
In view of this, for an $N\in\NN$, let $T_1,\ldots,T_N$ be a partitioning of $T$ into intervals $T_\ell=[T_\ell^-,T_\ell^+)$, that is, $T_1\cup\ldots \cup T_N=T$ and $T_{\ell}\cap T_{\ell'}=\emptyset$ for $\ell\neq\ell'$. For $\ell=1,\ldots, N$, define the symmetric functions $r_\ell$ over $(-t_{M-1},t_{M-1})$ as
$r_\ell(\delta)=\ind_{T_\ell}(|\delta|)$.
By this choice, for $r=r_\ell$, $\I_r$ takes the explicit form
\begin{equation}
\label{eq:haarintegral}
\I_{r_\ell}(t;d_\circ)=d_\circ^{-1}\big(\exp(-d_\circ  \tau_1)-\exp(-d_\circ \tau_2)\big),
\end{equation}
$\tau_1=\min \{\max \{T_\ell^-,0\} ,t\}$, $\tau_2=\max\{\min \{T_\ell^+,t\} ,0\}$.
Then, the corresponding $v_\ell(t)$ at the increasing sequence of times $t_k$ is efficiently found by numerical integration of $\eqref{eq:SigmaFrom0}$, with $Q=0$ and $V_\rho$ as in~\eqref{eq:Vstationary} with~\eqref{eq:haarintegral} in place of $\I_r$. Moreover, thanks to the piecewise constant nature of the $r_\ell$, positive semi-definiteness of $\R(\cdot)c$ is equivalent to positive semi-definiteness of the symmetric Toeplitz matrix with first row $[\,\R(t_0)  c,~\cdots,~\R(t_{M-1})  c\,]$, which we denote by $\T(c)$. 
Then, constraint~\eqref{eq:finitereconstructionconstraints} may be replaced by the more practical constraint
\begin{equation}
\label{eq:haarreconstructionconstraints}
\T(c) \in \C_M,
\end{equation}
with $\C_M$ the convex cone of symmetric, positive semi-definite matrices of order $M$.
We further define the penalty function
$$\Q(c)=\sum_{k=1}^{M-2} (\R(t_{k-1})  c-2  \R(t_{k})  c+\R(t_{k+1})  c)^2,$$
a natural adaptation of the standard second-order roughness penalty $\int(\ddot r(\delta))^2d\delta$ to a piecewise-constant function~\cite{DeNicolaoEtAl1997,Wahba,RasmussenWilliams}. 
The resulting problem can be straightforwardly implemented and solved by standard convex optimization software~\cite{CVX}.

In summary, the estimation procedure is as follows:
\begin{itemize}
 \item Given $\tilde\sigma^2_Y(\cdot)$, $\gamma$, and $T_1,\ldots,T_N$;
 \item Compute $v_0(t_k)$, with $k=0,\ldots, M-1$, by numerical integration of~\eqref{eq:SigmaFrom0} with $Q=Q^*$ and $V_\rho=0$;
 \item For $\ell=1,\ldots, N$ compute $v_\ell(t_k)$, with $k=0,\ldots, M-1$, by numerical integration of~\eqref{eq:SigmaFrom0} with $Q=0$ and using expressions~\eqref{eq:Vstationary} and~\eqref{eq:haarintegral} for $V_\rho$;
 \item Find a solution $\hat c$ to the optimization problem with objective function~\eqref{eq:finitereconstructionobjective} subject to constraint \eqref{eq:haarreconstructionconstraints};
 \item Return $\hat\rho_U(\cdot) =\R(\cdot)\hat c$.
\end{itemize}

Clearly, estimate $\hat\rho_U(\delta)$ will only be relevant for $|\delta|\in T$.
As an extension, the best choice of the regularization weight $\gamma$ could be made in several ways, using prior information about the expected regularity of $\bar\rho_U$, or via classical cross-validation techniques~(see~\cite{DeNicolaoEtAl1997} and references therein). For the latter, notice that repeated optimization over different candidate values of $\gamma$ does not require recomputation of the $v_\ell$, which turns out to be the computationally most expensive portion of the procedure. We will not pursue this point further. Feasibility of reconstruction and performance of the method are demonstrated on a numerical case study in the next section.

\section{Computational example: Promoter switching with stochastic rates}
\label{sec:example}

We now illustrate the theoretical results of Section~\ref{sec:momentequations} and~\ref{sec:spectralanalysis} in a simple reporter gene simulation study built upon the model of Section~\ref{sec:reportermodelling}. This will also serve as a numerical case study for the reconstruction method of Section~\ref{sec:reconstruction}. We focus on a binary promoter process $U$, and consider a case where the switching rates of $U$ are stochastic. Note that for such a system, if the mechanistic model of the switching rate laws is not known, parametric analysis and inference approaches are inapplicable. A model for the switching laws is introduced next for the sake of illustration and simulation. Knowledge of this model is of course not exploited in the application of our nonparametric analysis and reconstruction methods.
  
Suppose that the binding of some unknown transcriptional regulator is necessary for enabling activation of the promoter of interest. Let $B$ be the binary process that encodes the binding state of the transcriptional regulator at the promoter site. When the regulator is present, that is $B=1$, activation of the promoter, i.e. transition of $U$ from $0$ to $1$, is possible at a rate $\lambda_+$. When the regulator is absent, that is $B=0$, promoter activation is disabled, that is no switch of $U$ from $0$ to $1$ may occur. Overall, this gives a stochastic switch-on rate for $U$ equal to $B\cdot \lambda_+$.
Regardless of $B$, promoter deactivation, that is, transition of $U$ from $1$ to $0$, is possible at a switch-off rate $\lambda_-$. Binding and unbinding of $B$ are themselves modelled as random events occurring at rates $\beta_+$ and $\beta_-$, respectively. 

In order to explore the results of the previous sections, relating the statistics of $U$ with those of $Y$, simulations of the full gene expression process $(B,U,X)$ resulting from this regulatory mechanism can be easily carried out. Yet, since the joint process $(B,U)$ is a (continuous-time) finite Markov chain, the statistics of 
$U$ can also be determined analytically. 
Let
$p_t$ be the four-dimensional column vector whose $l$th entry is the probability that, at time $t$, $(B,U)$ is in the $l$th state of the list $\{(0,0),~(1,0),~(0,1),~(1,1)\}$.
Then $p_t=\exp(\Pi t)p_0$, where $\Pi$ is the transition kernel
$$\Pi=\begin{bmatrix}-\beta_+& \beta_-&\lambda_-& 0 \\ \beta_+ & -(\beta_-+\lambda_+)& 0 & \lambda_- \\ 0 & 0 & -(\lambda_-+\beta_+)& \beta_- \\ 0 & \lambda_+ & \beta_+ & -(\lambda_-+\beta_-) \end{bmatrix}.$$ 
\begin{proposition}
\label{thm:regulationstats}
Define $C_U=[\,0~0~1~1\,]^T$ and, for $\newt\geq t\geq 0$, $q_{\newt,t}^T=C_U^T\exp(\Pi(\newt-t))$.  It holds that
\begin{align*}
\mu_U(t)&=C_U^Tp_t, &\rho_U(\newt,t)&=q_{\newt,t}^T  (\emph{diag}(C_U)-p_tC_U^T)  p_t.
\end{align*}
For strictly positive $\beta_+$, $\beta_-$, $\lambda_+$ and $\lambda_-$, $U$ has the stationary statistics $\bar \mu_U=C_U^Tp_\infty$
and $$\bar \rho_U(\delta)=C_U^T\exp(\delta)  (\emph{diag}(C_U)-p_\infty C_U^T)  p_\infty,$$
where $p_\infty$ is the unique solution of $0=\Pi p_\infty$.
\end{proposition}

\begin{figure}
\centering
\subfigure[\label{fig:varfromcov}]{\includegraphics[width=\columnwidth]{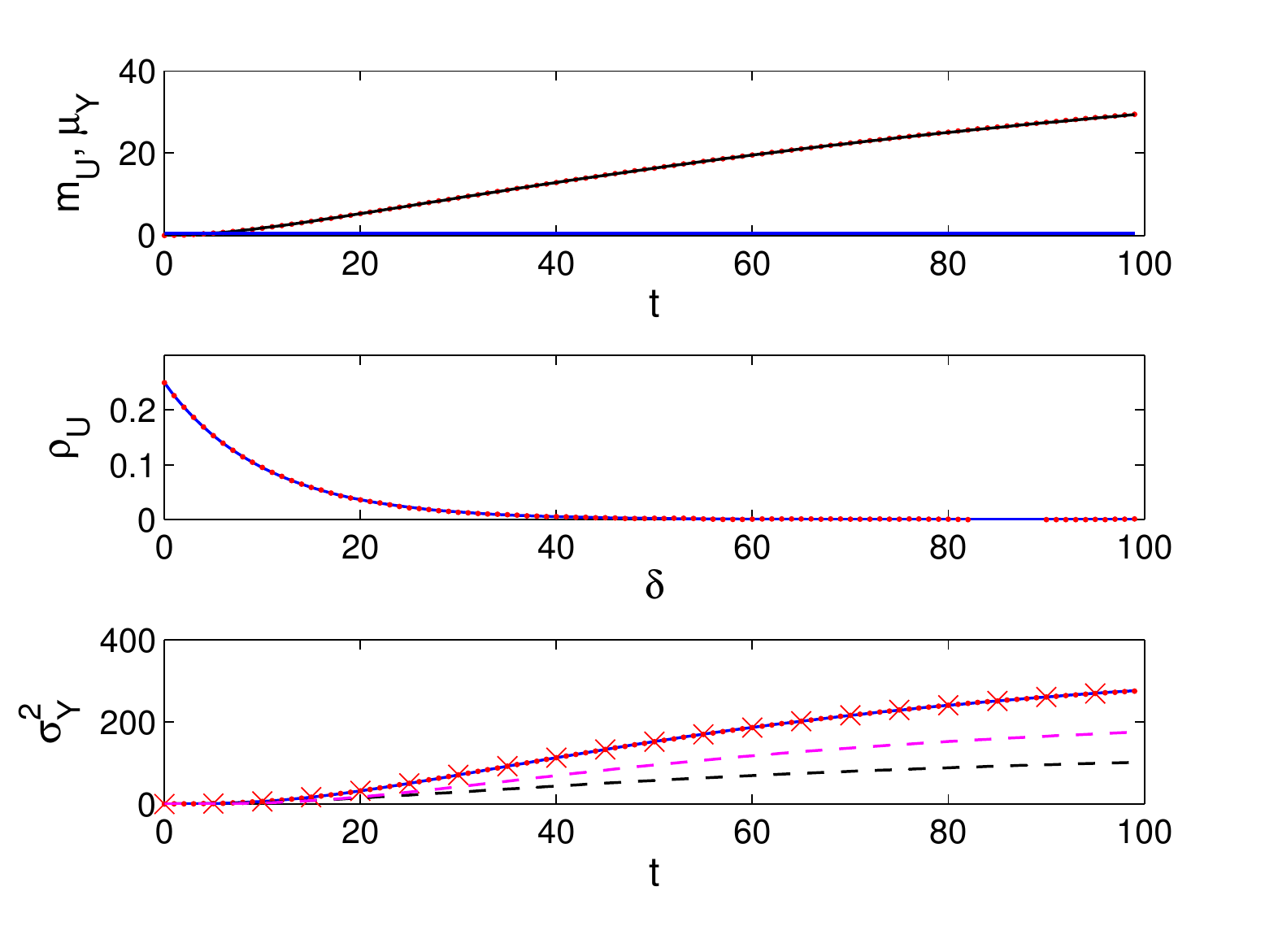}}\\
\subfigure[\label{fig:inoutspectra}]{\includegraphics[width=\columnwidth]{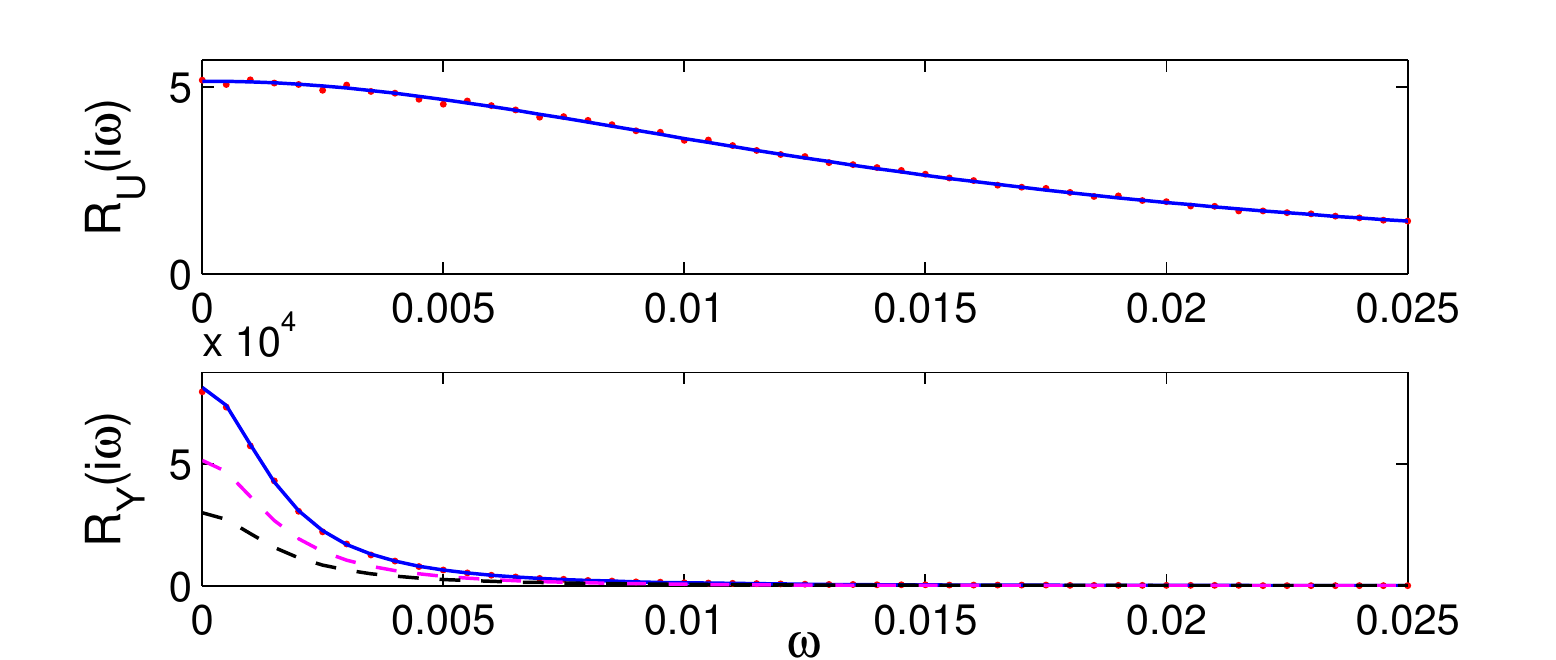}}
\caption{Statistics of the example process $U$ and the resulting process $Y$ (lines: analytical results; red dots: sample statistics of input $U$ and output $Y$ from $10^5$ simulations of the stochastic system; red crosses: measurements utilized in a single run of the numerical study of the covariance reconstruction problem). \subref{fig:varfromcov} 
Top: Stationary input mean $\bar\mu_U$ (blue) and nonstationary output mean $\mu_Y(\cdot)$ (black); Center: Stationary input autocovariance $\bar\rho_U(\cdot)$; Bottom: Nonstationary output variance $\sigma_Y^2(\cdot)$ (blue), and its decomposition in a component independent of $\bar \mu_U$ (dashed magenta) and a component depending on $\bar \mu_U$ (dashed black). \subref{fig:inoutspectra} Top: Input spectrum $R_U(i\omega)$; Bottom: Output spectrum $R_Y(i\omega)$ when $Y$ has reached stationarity, and its decomposition into $H(i\omega)H(-i\omega)R_U(i\omega)$ (dashed magenta) and $R_E(i\omega)$ (dashed black). Empirical spectra are computed from $10^4$ simulated trajectories reaching stationarity. Time unit is the minute. The dashed-black component in~\subref{fig:varfromcov}(bottom) is equal to $v_0(\cdot)$ (see Section~\ref{sec:reconstruction}) and is associated with the same dynamics that give rise to the dashed-black spectral component in~\subref{fig:inoutspectra}(bottom).
}
\label{fig:demo} 
\end{figure}

On the basis of this, Fig.~\ref{fig:demo} illustrates the results of the previous sections on the computation of the statistics of $Y$ as a function of those of $U$, for the case where $U$ is in stationary conditions. Reporter parameters were fixed to biologically relevant values inspired by~\cite{KaernEtAl2005}, $\theta=(0.5, 0.1, 0.2, 0.01)$ [min$^{-1}$]. Switching process parameters $(\lambda_+,\lambda_-)=(0.1,0.05)$ [min$^{-1}$] and $(\beta_+,\beta_-)=(0.1351,0.1)$ [min$^{-1}$] are such that $U$ spends on average an equal time in the on and off states (that is, $\bar\mu_U=0.5$ by construction).
For comparison, numerical Monte Carlo simulations of process $(B,U,X)$ were carried out in StochKit~\cite{StochKit2}. The initial conditions $\big(B(0),U(0)\big)$ over the $10^5$ simulations were assigned in proportion to $p_\infty$ to obtain sampling of $(B,U)$ in stationary conditions, while 
the initial condition of $X$ was fixed to $X(0)=0$. The empirical statistics drawn from these simulations are also reported in the figure and are found to be in perfect agreement with the analytic developments.

We now demonstrate the autocovariance reconstruction procedure of Section~\ref{sec:reconstruction} on the example introduced above. In stationary conditions, the statistics of process $U$ are illustrated in Fig.~\ref{fig:demo}. As illustrated in the bottom plot of Fig.~\ref{fig:varfromcov}, we consider the case where 
measurements $\tilde \sigma_Y^2$ are taken at $M=20$ time instants $t_k=5\cdot k$ [min], with $k=0,\ldots, M-1$, from independent simulated samples of $10^5$ cells, based on the random simulation of $(B,U,X)$. Sparsity and size of the data set are relevant to real-world experimental scenarios.
The family of basis functions is defined by $N=96$, $T_1=[0,0.5)$, $T_N=[94.5,95)$ and, for $\ell=2,\ldots,N-1$,  $T_\ell=[\ell-1.5,\ell-0.5)$. This essentially corresponds to a uniform partitioning of the measurement period $T=[0,95]$. Notice that $N\gg M$ makes regularization crucial. 
Estimation of $\hat\rho_U$ is performed over $1000$ randomly generated datasets, yielding $1000$ estimates that are used to compute empirical mean and standard deviation of $\hat \rho_U(\delta)$, with $\delta\in T$.

\begin{figure}
\centering
\includegraphics[width=\columnwidth]{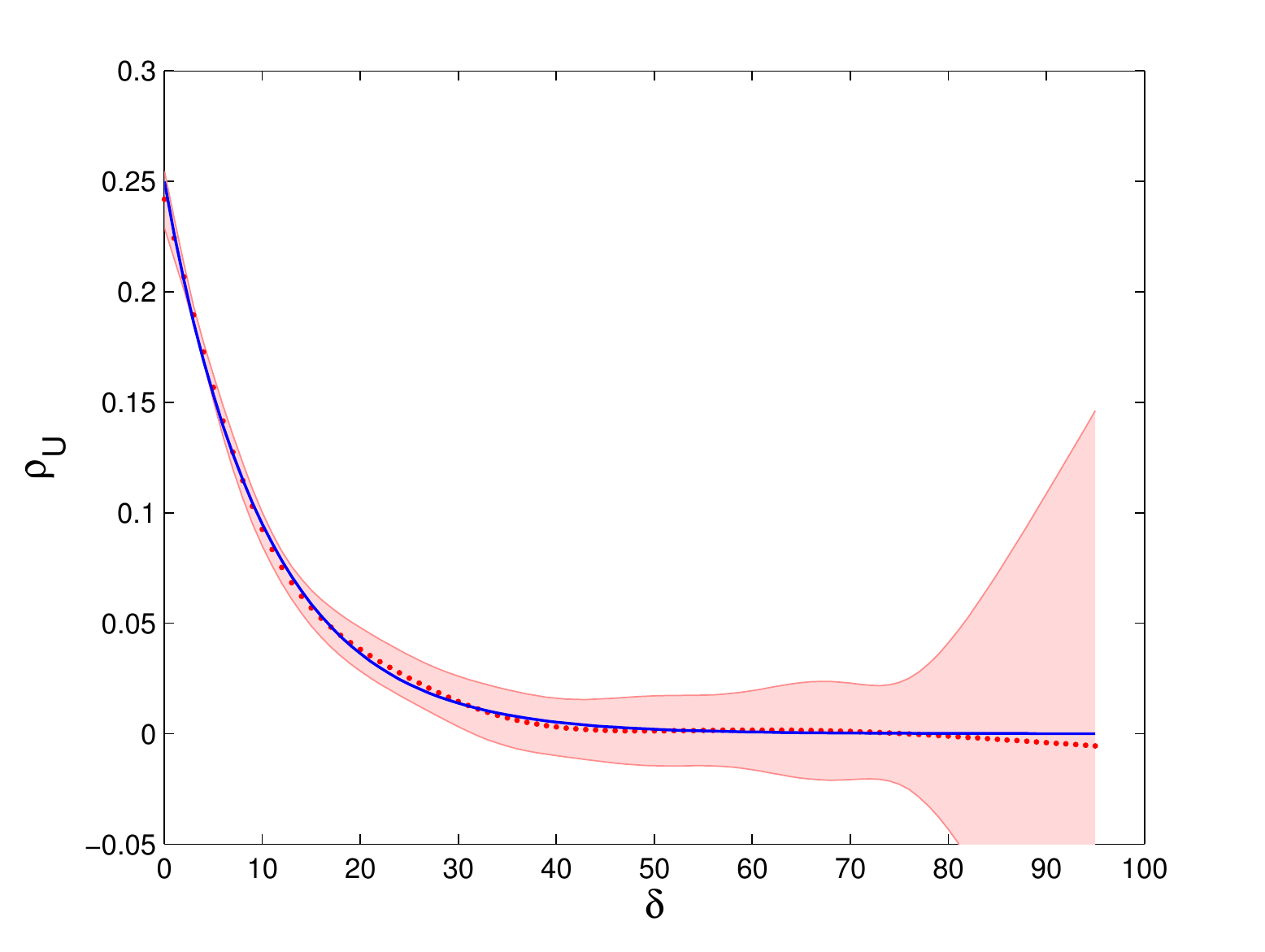}
\caption{Statistics of the estimation of $\bar \rho_U(\delta)$ (blue line) at lags $\delta=0,1,\ldots, 95$ ($\gamma=10^5$). Red dots: $\mean\big(\hat\rho_U(\cdot)\big)$ from $1000$ simulated runs; Shaded red region: Confidence band defined by $\mean\big(\hat\rho_U(\cdot)\big) \pm 2\cdot\std\big(\hat\rho_U(\cdot)\big)$ (bottom-cropped for better scaling).}\label{fig:reconstruction}
\end{figure}

The performance of estimation is illustrated in Fig.~\ref{fig:reconstruction}. The first observation is that the true covariance function is, as desired, contained in the confidence band $\mean\big(\hat \rho_U(\delta)\big)\pm 2\cdot \std\big(\hat \rho_U(\delta)\big)$, $\delta\in T$. This band is remarkably narrow for small lags and is
nicely bounded for $\delta$ roughly up to $75$. Larger lags are instead affected by diverging uncertainty. This is a clear sign of the little sensitivity of the measurements to the tails of $\bar\rho_U(\delta)$, that can be understood in terms of the asymptotic stability of the generalized moment equations, or in other words, the decaying memory of the reporter process. A second observation is that, despite regularization, estimates show very limited bias. For the case into study, because $U$ is a Bernoulli process, the convex constraint $\bar \mu_U^*(1-\bar \mu_U^*)=\bar \rho_U(0)$, relating the mean (assumed known) with the variance $\var(U)=\bar \rho_U(0)$ of the process, could also be added into the optimization problem to ameliorate estimation around $\delta=0$. From the results obtained, though, this seems superfluous.

\section{Conclusions}
\label{sec:conclusions}

In this paper we have addressed analysis of first-order stochastic reaction kinetics with an input process. We have derived analytic relationships between first- and second-order statistics of the input process and those of the random network state. In stationary conditions, we have provided a spectral characterization of these networks and discussed connections with spectral analysis of linear dynamical systems with stochastic inputs. We have specialized the results to the case of the gene reporter systems commonly utilized in biology, and derived an interpretation of these sensoring systems in terms of linear noisy filters. Then, we have applied the theoretical results to derive a method for nonparametric reconstruction of a stationary gene promoter autocovariance function from transient population snapshot data. On a simulated gene expression case study, we have demonstrated the theoretical results and assessed feasibility of reconstruction and performance of the method.

Results of this paper are of immediate interest for the investigation of gene expression regulation, since autocovariance reconstruction allows one to learn about gene expression time scales and memory. Because our approach is nonparametric, no \textit{a priori} assumption on the mechanisms of promoter regulation are needed. Rather, in the perspective of gene network analysis, our approach should be considered as a first step decoupled from and enabling the subsequent exploration of regulatory interactions. 
In this perspective, our results are in the same spirit of~\cite{MunskyEtAl2009,LNCS2009}, where stochastic analysis is leveraged to outperform deterministic approaches in reconstruction of unknown network parameters. Yet the interest of our results goes beyond gene expression analysis. The theoretical results presented in the paper apply to any stochastic reaction network with state-affine rates or that can be approximated as such. Besides their natural relevance for biology and chemistry, the analysis and suitable extensions of the methods proposed are of more general interest to applications based on CTMC models in particular~\cite{NorrisBook}, and stochastic hybrid systems in general~\cite{LygerosPrandini}.

A number of developments of our work are envisioned. Concerning biological applications, the input autocovariance reconstruction method has been developed under a few assumptions, notably input stationarity and determinism of the initial network state. In order to broaden applicability, it is of interest to relax these assumptions. More generally, thanks to linearity, the differential equations relating input and output second-order moments can be employed to address a variety of inverse problems very conveniently, for instance, the reconstruction of a nonstationary input autocovariance from readouts of the reporter autocovariance function. This problem resonates challenges from other fields, for instance, geostatistics~\cite{Chiles}. Whereas population snapshot data do not provide time correlations, other experimental techniques, such as fluorescence videomicroscopy, do so and enable in principle to tackle the problem. In all these problems, data quality of course determines achievable performance. As a further application of the work presented, spectral analysis of reporter systems can be utilized for their optimal design, such that the error component brought about by the stochasticity of the network is minimized relative to the information conveyed about the input process. Generalization of our results to more complex reaction rate models would also be of interest. While a time-varying deterministic expression for the rate matrix $W$ is easily accommodated in most of our results, nonlinear dependence on the state $X$, such as for second-order reactions, should be handled via approximations~\cite{RuessEtAl2011,KlimovskaiaEtAl,GillespieMomentClosure,SinghHespanhaFeb11}. 

From a system-theoretic perspective, fast-developing experimental biotechnologies provide increasing insight into the behavior of individual cells, but also open up new challenges in data processing and model reconstruction. Optimal exploitation of the data requires new methodologies that can profit from the large bulk of knowledge developed in the automatic control and signal processing communities. The peculiarities of biological experimentation make these problems nonstandard though, such that original challenges coming from these applications will keep providing new feed to the estimation and identification research fields.

\section*{Acknowledgements}
This work was funded in part by the French national research agency (ANR) via project MEMIP (ANR-16-CE33-0018). The author thanks Hidde de Jong for useful suggestions in improving the manuscript.

\bibliographystyle{plain}      
\bibliography{GMM}

\appendix

\section{Proofs}
\label{asec:proofs}

\paragraph*{Proof of Proposition~\ref{thm:conditionalmoments}.}

The differential equations for $\mu^{f,x_0}$, $\Sigma^{f,x_0}$ and $\rho^{f,x_0}$ are available in the literature~\cite{Hespanha,LestasEtAl2008}.
The explicit solutions of the differential equations for the mean $\mu^{f,x_0}$, relative to the initial condition $\mu^{f,x_0}(0)=x_0$, and for the autocovariance matrix $\rho^{f,x_0}$, relative to the initial condition $\rho^{f,x_0}(t)=\Sigma^{f,x_0}(t)$, are apparent. The solution for $\Sigma^{f,x_0}$ follows from the general solution of matrix Lyapunov equations of the form 
$\dot \Sigma(t)=A\Sigma(t)+\Sigma(t)A^T+Q(t)$
which is 
$\Sigma(t)=\exp(At)\Sigma(0)\exp(At)^T+
\int_0^t \exp\big(A(t-\tau\big)Q(\tau)\exp\big(A(t-\tau\big)^Td\tau.
$
In the present case, $A=SW$ and $Q(t)=Q^{f,x_0}(t)$ is as defined in the statement, while $\Sigma(0)=\Sigma^{f,x_0}(0)=0$. The results of the proposition thus follow from the definition of $\ell(t)$ and of the integral operators $\Lscr_t$ and $\Lscr_t^\circ$.

\paragraph*{Proof of Proposition~\ref{thm:moments}.}

The results follow from marginalization of the conditional statistics computed in Proposition~\ref{thm:conditionalmoments} with respect to the conditioning process $F$ and initial condition $X_0$, also using commutation of expectation with integration~\cite{Jazwinski1970}. For the mean, 
$\mu(t)=\EE[\mu^{F,X0}(t)]=\EE[\ell(t)  X(0)+\Lscr_t(S   F)]=\ell(t)\mu_0+\Lscr_t(S \mu_F)$.
To compute $\rho(\newt,t)$ for $\newt\geq t$, the rationale is to marginalize the centered moments 
$M(\newt,t)^{F,X_0}\triangleq\EE[X(\newt)X(t)^T|F,X_0]=\rho^{F,X_0}(\newt,t)+\mu^{F,X_0}(\newt)\mu^{F,X_0}(\newt)^T$ to obtain $M(\newt,t)\triangleq\EE[X(\newt)X(t)^T]$, and then infer the autocovariance from $\rho(\newt,t)=M(\newt,t)-\mu(\newt)\mu(t)^T$. We will first express $\rho(\newt,t)$ as a function of $\Sigma(t)$, then work out the the expression of latter, and finally merge the results. 
To express $\rho(\newt,t)$ as a function of $\Sigma(t)$, note that one may also write $\mu^{f,x_0}(\newt)=\ell(\newt-t)\mu^{f,x_0}(t)+\Lscr_{t,\newt}(S f)$ and $\mu(\newt)=\ell(\newt-t)\mu(t)+\Lscr_{t,\newt}(S \mu_F)$, where, for any relevant $g$,
$\Lscr_{t,\newt}g=\int_t^\newt d\tau~\ell(\newt-\tau)g(\tau)$. Then, also using Proposition~\ref{thm:conditionalmoments},
\begin{align*}
M(\newt,t)&=\EE\left[M(\newt,t)^{F,X_0}\right] \\
&=\EE\left[\rho^{F,X_0}(\newt,t)\right]+\EE[\mu^{F,X_0}(\newt)\mu^{F,X_0}(t)^T] \\
&=\EE\left[\ell(\newt-t)\Sigma^{F,X_0}(t)\right]+\EE\left[\big(\ell(\newt-t)\mu^{F,X_0}(t)+\right. \\
&\qquad \left.\Lscr_{t,\newt}(S   F)\big) \mu^{F,X_0}(t)^T\right] \\
&=\ell(\newt-t)\EE\left[\Sigma^{F,X_0}(t)+\mu^{F,X_0}(t) \mu^{F,X_0}(t)^T\right] +\\
&\qquad \EE\left[\Lscr_{t,\newt}\left(S   F\right)  \mu^{F,X_0}(t)^T\right] \\
&=\ell(\newt-t)M(t)+\Lscr_{t,\newt}\left(S  \EE\left[F  \mu^{F,X_0}(t)^T\right]\right),\\
\rho(\newt,t)&=M(\newt,t)-\mu(\newt)\mu(t)^T \\
&=\ell(\newt-t)M(t)+\Lscr_{t,\newt}\left(S  \EE\left[F  \mu^{F,X_0}(t)^T\right]\right)- \\
&\qquad \left(\ell(\newt-t)\mu(t)+\Lscr_{t,\newt}(S \mu_F) \right)  \mu(t)^T \\
&=\ell(\newt-t) \left(M(t)-\mu(t)\mu(t)^T\right)+\\
&\qquad \Lscr_{t,\newt}\left(S  \EE\left[F\mu^{F,X_0}(t)^T\right]-S \cdot \mu_F  \mu(t)^T\right) \\
&=\ell(\newt-t) \Sigma(t)+\Lscr_{t,\newt}\left(S \Cov\big(F,\mu^{F,X_0}(t)\big) \right).
\end{align*}
The rightmost term of the last line can in turn be expanded as
$\Lscr_{t,\newt}\left(S   \Cov\big(F,\ell(t)  X(0)+\Lscr_t(S   F)\big) \right)=
\Lscr_{t,\newt}(S   \xi_F^T)  \ell(t)^T+\Lscr_{t,\newt}\Lscr_t^*\left(S   \rho_F  S^T \right)
$,
where we have used the fact that, for the generic $g$, $(\Lscr g)^T=\Lscr^*(g^T)$, and exchanged expectation with integration.
We next focus on the computation of $\Sigma(t)$, which we perform via that of
$M(t)$. Using Proposition~\ref{thm:conditionalmoments}, one gets
\begin{align*}
M(t)&=\EE\left[\Sigma^{F,X_0}(t)\right]+\EE[\mu^{F,X_0}(t)\mu^{F,X_0}(t)^T] \\
&=\EE\left[ \Lscr^\circ_t \left(S\diag\big(W \mu^{F,X_0}+ F\big)S^T\right)\right]+\\
&\qquad \EE\left[\left(\ell(t) X(0)+\Lscr_t(S  F)\right) \left(\ldots\right)^T\right] \\
&= \Lscr^\circ_t \left(S\diag\big(W \mu+ \mu_F\big)S^T\right)+\ell(t)M(0)\ell(t)^T+\\
&\qquad \EE\left[\Lscr_t(S  F)\Lscr_t^*(F^T S^T)\right]+\ell(t)\EE\left[X(0)\times\right. \\
&\qquad \left. \Lscr_t^*(F^T S^T)\right]+\EE\left[\Lscr_t(S F) X(0)^T\right]\ell(t)^T.
\end{align*}
Substituting the above and the expression of $\mu(t)$ into $\Sigma(t)=M(t)-\mu(t)\mu(t)^T$, regrouping terms and using 
$M(0)-\mu_0\mu_0^T=\Sigma_0$, one gets that $\Sigma(t)$ is equal to
\begin{multline*}
\Lscr^\circ_t \left(S\diag\big(W  \mu+ \mu_F\big)S^T\right)+\ell(t)\Sigma_0\ell(t)^T+\\
\left(\EE\left[\Lscr_t(S  F)\Lscr_t^*(F^TS^T)\right]-\Lscr_t(S \mu_F)
\Lscr_t^*(\mu_F^T S^T) \right)+ \\
\ell(t)\left(\EE\left[X(0)\Lscr_t^*(F^T  S^T)\right]-
\mu_0\Lscr_t^*(\mu_F^TS^T)\right)+\\
\left(\EE\left[\Lscr_t(S  F)X(0)^T\right]-\Lscr_t(S \mu_F)\mu_0^T\right)\ell(t)^T\\
=\Lscr^\circ_t \left(S\diag\big(W  \mu+ \mu_F\big)S^T\right)+\ell(t)\Sigma_0\ell(t)^T+\\
\Lscr_t\Lscr_t^*(S \rho_F S^T)+
 \ell(t) \Lscr_t^*\left(\xi_F S^T\right)+\Lscr_t\left(S \xi_F^T\right) \ell(t)^T
\end{multline*}
where, for the relevant $g$ and $g'$, we have used the fact that $\EE[(\Lscr g) (\Lscr^*g')]=\EE[\Lscr\Lscr^*(gg')]=\Lscr\Lscr^*(\EE[gg'])$. 
Finally, plugging this into the expression of $\rho(\newt,t)$ above, using the equality $\ell(\newt-t)\ell(t)=\ell(\newt)$ and regrouping terms, one gets that $\rho(\newt,t)$ is equal to
\begin{multline*}
\ell(\newt-t)  \Lscr^\circ_t \left(S\diag\big(W  \mu+ \mu_F\big)S^T\right)+
\ell(\newt)\Sigma(0)\ell(t)^T+ \\
\left(\ell(\newt-t)\Lscr_t\Lscr_t^*(S  \rho_F  S^T)+\Lscr_{t,\newt}\Lscr_t^*\left(S   \rho_F  S^T \right)\right)+ \\
\left(\ell(\newt-t)\Lscr_t\left(S  \xi_F^T\right)  \ell(t)^T+
\Lscr_{t,\newt}\left(S   \xi_F^T  \ell(t)^T\right)\right)+
\\
\ell(\newt) \Lscr_t^*\left(\xi_F  S^T\right)\\
=\ell(\newt-t) \Lscr^\circ_t \left(S\diag\big(W  \mu+ \mu_F\big)S^T\right)+
\ell(\newt)\Sigma(0)\ell(t)^T+\\
\Lscr_\newt\Lscr_t^*(S \rho_F S^T)+
\Lscr_\newt\left(S \xi_F^T\right) \ell(t)^T+\ell(\newt) \Lscr_t^*\left(\xi_F S^T\right),
\end{multline*}
where, thanks to Fubini's theorem, $\Lscr_\newt$ and $\Lscr_t^*$ may be exchanged as in the statement of the result.

\paragraph*{Proof of Proposition~\ref{thm:stationarymoments}.}

To find the expression of the stationary mean $\bar\mu$, the easiest way is to recognize that the expression of $\mu(t)$ in
Proposition~\ref{thm:moments} is a solution of the differential equation
$\dot \mu = SW\mu+S \mu_F$.
For $SW$ strictly stable and constant $\bar \mu_F$, regardless of initial conditions, the asymptotically stable equilibrium $\bar\mu$ is the solution in $\mu$ of $0=\dot \mu=SW  \mu+S \bar \mu_F$. Since $SW$ is invertible, the expression of $\bar\mu$ follows. For the stationary autocovariance matrix function $\bar\rho(\cdot)$, first notice that the elements of $\xi_F$ are uniformly bounded thanks to the uniform boundedness of the second-order moments of $F$, and that, for $t\to +\infty$, $\ell(t)\to 0$ thanks to strict stability of $SW$. Then, from the expression of $\rho(\newt,t)=\rho(t+\delta,t)$ in Proposition~\ref{thm:moments},
\begin{align*}
\bar\rho(\delta)&=\lim_{t\to+\infty} \Lscr^*_t\Lscr_{t+\delta}(S   \rho_F  S^T)+ \\
&\qquad\ell(\delta)\lim_{t\to+\infty} \Lscr^\circ_t\left(S\diag\big(W\mu+ \mu_F\big)S^T\right),
\end{align*}
where we have already eliminated the vanishing term $\ell(t+\delta)\Sigma_0\ell(t)^T$ and terms 
$\Lscr_{t+\delta}\left(S  \xi_F^T\right)  \ell(t)^T$ and $\ell(t+\delta)  \Lscr_t^*\left(\xi_F  S^T\right)$,
which also vanish since $\Lscr\left(S  \xi_F^T\right)$ and $\Lscr^*\left(\xi_F  S^T\right)$ are themselves bounded (they are both integrals of an exponentially stable kernel times a uniformly bounded factor). We are thus left with the computation of the two limits above. 
For the first, under the assumption that $\rho_F(t+\delta,t)=\bar\rho_F(\delta)$, note that $\Lscr^*_t\Lscr_{t+\delta}(S   \rho_F  S^T)$ is equal to
\begin{multline*}
\int_0^{+\infty}d\zeta\int_0^{+\infty}d\tau~\ell(t+\delta-\tau)S   \bar \rho_F(\tau-\zeta)  S^T \ell(t-\zeta)^T \\
=\int_{-\infty}^t d\zeta'\int_{-\infty}^{t+\delta} d\tau'~\ell(\tau')S   \bar \rho_F(\delta+\zeta'-\tau')  S^T \ell(\zeta')^T
\end{multline*}
where we have made use of the change of variables $\tau'=t+\delta-\tau$, $\zeta'=t-\zeta$. Thus, in the limit, one gets
\begin{multline*}
\int_{-\infty}^{+\infty} d\zeta'\int_{-\infty}^{+\infty} d\tau'~\ell(\tau')S \bar \rho_F(\delta+\zeta'-\tau') S^T \ell(\zeta')^T=\\
\int_{-\infty}^{+\infty} d\zeta' [\ell\conv S  \bar\rho_F](\delta+\zeta') S^T \ell(\zeta')^T= \\
\int_{-\infty}^{+\infty} d\zeta'' [\ell\conv S   \bar\rho_F](\delta-\zeta'') S^T \ell_-(\zeta'')^T=\\
\big[(\ell\conv S  \bar\rho_F  S^T)\conv \ell_-^T\big] (\delta),
\end{multline*}
where we have made use of the new change of variables $\zeta''=-\zeta'$, and the final writing may be simplified in view of the associativity of convolution.

For the second, given $\mu_F(t)=\bar \mu_F$, it suffices to show that the limit is equal to that of
$\Lscr^\circ_t\left(S\diag\big(W\bar \mu+\bar\mu_F\big)S^T\right)$.
Indeed, $\Lscr^\circ_t\left(S\diag\big(W\mu+\bar\mu_F\big)S^T\right)$
is the solution at $t$ of the system of ODEs
\begin{align*}
\dot \mu &= SW \mu+S \bar\mu_F \\
\dot \Sigma^\circ&=SW \Sigma^\circ + \Sigma^\circ W^TS^T + S\diag\big(W  \mu+ \bar\mu_F\big)S^T
\end{align*}
with the initial condition $\Sigma^\circ(0)=0$. Since $SW$ is strictly stable, for $t\to+\infty$, the system converges to the unique equilibrium given by $\mu=\bar\mu$ and $\Sigma^\circ=\bar \Sigma^\circ$, for which
$$0=SW \bar \Sigma^\circ + \bar \Sigma^\circ W^TS^T + S\diag\big(W  \bar \mu+\bar\mu_F\big)S^T.$$
In turn, this is also the unique stationary solution of 
$$\dot \Sigma^\circ=SW \Sigma^\circ + \Sigma^\circ W^TS^T + S\diag\big(W  \bar \mu+ \bar\mu_F\big)S^T$$
with $\Sigma^\circ(0)=0$, whose solution at time $t$ is precisely $\Lscr^\circ_t\left(S\diag\big(W  \bar \mu+ \bar\mu_F\big)S^T\right)$, which is what was sought. It only remains to notice that the limit of $\Lscr^\circ_t\left(S\diag\big(W  \bar \mu+\bar\mu_F\big)S^T\right)$ is equal to
$$
\Lscr^\circ_\infty\left(S\diag\big(-W(SW)^{-1}S \bar\mu_F+ \bar\mu_F\big)S^T\right)= \\
\Lscr^\circ_\infty(\bar Q),
$$
where the expression of $\bar \mu$ as a function of $\bar\mu_F$ has been inserted.

\paragraph*{Proof of Proposition~\ref{thm:reporterspectrum}.}

For the definition of $S$, $W$ and $F$ in Section~\ref{sec:reportermodelling}, we need to compute the second-row, second-column element of~\eqref{eq:spectrum}.
To do this we start by computing the expression of $\bar \Sigma^\circ$ as the solution of~\eqref{eq:sigmacirc}. In turn, this entails the computation of $\bar\mu_F$ and $\bar Q$. From the definition of $F$, it is immediately found that $\bar\mu_F=[\,k_M \bar \mu_U~ 0~ 0~ 0\,]^T$. In view of the zero entries of $\bar\mu_F$, by the definition of $\bar Q$, it suffices to compute the first column of $I-W(SW)^{-1}S$, which is promptly found to be $[\,1~1~k_P/d_M~k_P/d_M\,]^T$. Thus
$\bar Q = k_M \bar \mu_U  S \diag(1,1,k_P/d_M,k_P/d_M)  S^T=2k_M \bar \mu_U \diag(1,k_P/d_M)$. Using this, by the symmetry of $\bar \Sigma^\circ$, the solution of~\eqref{eq:sigmacirc} is easily written in terms of a system of three linear equations (two for the diagonal entries and one for the off-diagonal term). The solution of this yields
$\bar\Sigma^\circ=\begin{bmatrix} 1 & \alpha \\ \alpha & r_P (1+\alpha) \end{bmatrix}  r_M \bar \mu_U$.
To get to the result of the proposition, it remains to find expressions for $R_F(i\omega)$ and $L(i\omega)$. Again by the definition of $F$, the first is the Fourier transform of
$\diag(k_M^2\bar \rho_U(\delta),0,0,0)$, that is $R_F(i\omega)=\diag(k_M^2R_U(i\omega),0,0,0)$. The second is found via the Laplace transform of $\ell(t)$, the latter being the impulse response of a linear dynamical system with state matrix $SW$. Therefore 
$$L(\sl)=(\sl I-SW)^{-1}=\begin{bmatrix}\frac{1}{d_M+\sl} & 0 \\ \frac{k_P}{(d_M+\sl)(d_P+\sl)} & \frac{1}{d_P+\sl}\end{bmatrix}.$$
Finally, inserting the expressions of all factors in~\eqref{eq:spectrum}, the result follows after basic algebra.

\paragraph*{Proof of Proposition~\ref{thm:regulationstats}.}

Denote with $\vind_l$ the four-dimensional column vector whose entries are zero except for the entry in the $l$th row taking value $1$, with $l=1,\ldots, 4$.
Note that $\vind_3^Tp_t=\PP[B(t)=0,U(t)=1]$ and $\vind_4^Tp_t=\PP[B(t)=1,U(t)=1]$.  
Thus
$\mu_U(t)=\PP[U(t)=1]=\vind_3^Tp_t+\vind_4^Tp_t=C_U^T  p_t$.
To compute $\rho_U(\newt,t)$, with $\newt\geq t$, we use the relation $p_\newt=\exp\big(\Pi(\newt-t)\big)p_t$ to first compute $\EE[U(\newt)U(t)]=\PP[U(\newt)=1,U(t)=1]$ as
\begin{multline*}
\PP[U(\newt)=1|U(t)=1,B(t)=0]\cdot \PP[U(t)=1,B(t)=0] \\
+\PP[U(\newt)=1|U(t)=1,B(t)=1]\cdot \PP[U(t)=1,B(t)=1] 
\end{multline*}
which is equal to (recall $\delta=\newt-t$)
\begin{multline*}
\left( C_U\exp(\Pi\delta)\vind_3\right) \left(\vind_3^T p_t\right)+\left(C_U\exp(\Pi\delta)\vind_4\right) \left(\vind_4^T p_t\right)=\\
 C_U\exp(\Pi\delta)\left(\vind_3\vind_3^T+\vind_4\vind_4^T\right) p_t=
q_{\newt,t}^T\diag(C_U)p_t.
\end{multline*}
Then, using $\mu_U(\newt)=C_U^Tp_\newt=q_{\newt,t}^Tp_t$, $\rho_U(\newt,t)$ is given by
\begin{multline*}
\EE[U(\newt)U(t)]-\mu_U(\newt)\mu_U(t)=q_{\newt,t}^T\diag(C_U)p_t-\\
(q_{\newt,t}^Tp_t)(C_U^Tp_t)=q_{\newt,t}^T(\diag(C_U)-p_tC_U^T)p_t.
\end{multline*}
Next, it is easily verified that, for strictly positive parameters $\beta_+$, $\beta_-$, $\lambda_+$ and $\lambda_-$, the chain is irreducible and recurrent. Then, the stationary distribution $p_\infty$ satisfying $\Pi p_\infty$ exists, is unique and is also the limiting distribution $p_t$ as $t\to+\infty$, irrespective of $p_0$
The stationary statistics are thus found by replacing $\newt$ with $t+\delta$ in the above formulas and taking the limit for $t\to+\infty$.

\end{document}